\documentclass[11pt]{article}
\usepackage{amsfonts}

\usepackage[preprint]{acl}
\usepackage{amsmath}
\usepackage{times}
\usepackage{latexsym}
\usepackage{multirow}
\usepackage{hyperref}

\usepackage{placeins}  
\usepackage{algorithm}
\usepackage{subcaption}

\usepackage{algorithmic}
\usepackage{multicol}
\usepackage{booktabs}   
\usepackage{colortbl}   
\usepackage[T1]{fontenc}

\usepackage[utf8]{inputenc}

\usepackage{microtype}

\usepackage{inconsolata}

\usepackage{graphicx}

\usepackage{comment}
\def\figvsp{\vspace{0cm}}
\def\calm/{CALM}
\title{CALM: Class-Conditional Sparse Attention Vectors for \\
Large Audio-Language Models} 

\author{Videet Mehta \\
  MIT CSAIL \\
  \texttt{mvideet@mit.edu} \\\And
  Liming Wang \\
  MIT CSAIL \\
  \texttt{limingw@mit.edu} \\\And
  Hilde Kuehne \\
  Tuebingen AI Center \\
  MIT-IBM Watson AI Lab\\
  \texttt{kuehne@cs.uni-bonn.de} \\\AND
  Rogerio Feris \\
  MIT-IBM Watson AI Lab \\
  \texttt{rsferis@us.ibm.com} \\\And
  James R. Glass \\
  MIT CSAIL \\
  \texttt{glass@mit.edu} \\\And
  M. Jehanzeb Mirza \\
  MIT CSAIL \\
  \texttt{jmirza@mit.edu}
}

\begin{document}
\maketitle

\begin{abstract}
Large audio-language models (LALMs) show strong zero-shot performance on tasks such as audio question answering, but still lag behind specialized systems on discriminative tasks such as audio classification. Recent work shows that sparse subsets of attention heads can serve as discriminative feature extractors for few-shot classification, but existing methods weight all selected heads equally. We propose \textbf{C}lass-Conditional Sparse \textbf{A}ttention Vectors for Large Audio-\textbf{L}anguage \textbf{M}odels (\calm/), a few-shot classifier that learns class-conditional importance weights over attention heads. By treating heads as class-specific experts and aggregating them with reliability-weighted soft voting, \calm/ allows different classes to rely on different subsets of heads. Across diverse few-shot audio, audiovisual, visual, and spoofing benchmarks, \calm/ consistently outperforms state-of-the-art uniform-voting baselines, with gains of up to 14.52, 1.53, and 8.35 absolute points on audio, audiovisual, and spoofing tasks, respectively. Code repository will be released upon acceptance.

\end{abstract}
\begin{figure}[t!]
    \centering
    \includegraphics[width=\columnwidth]{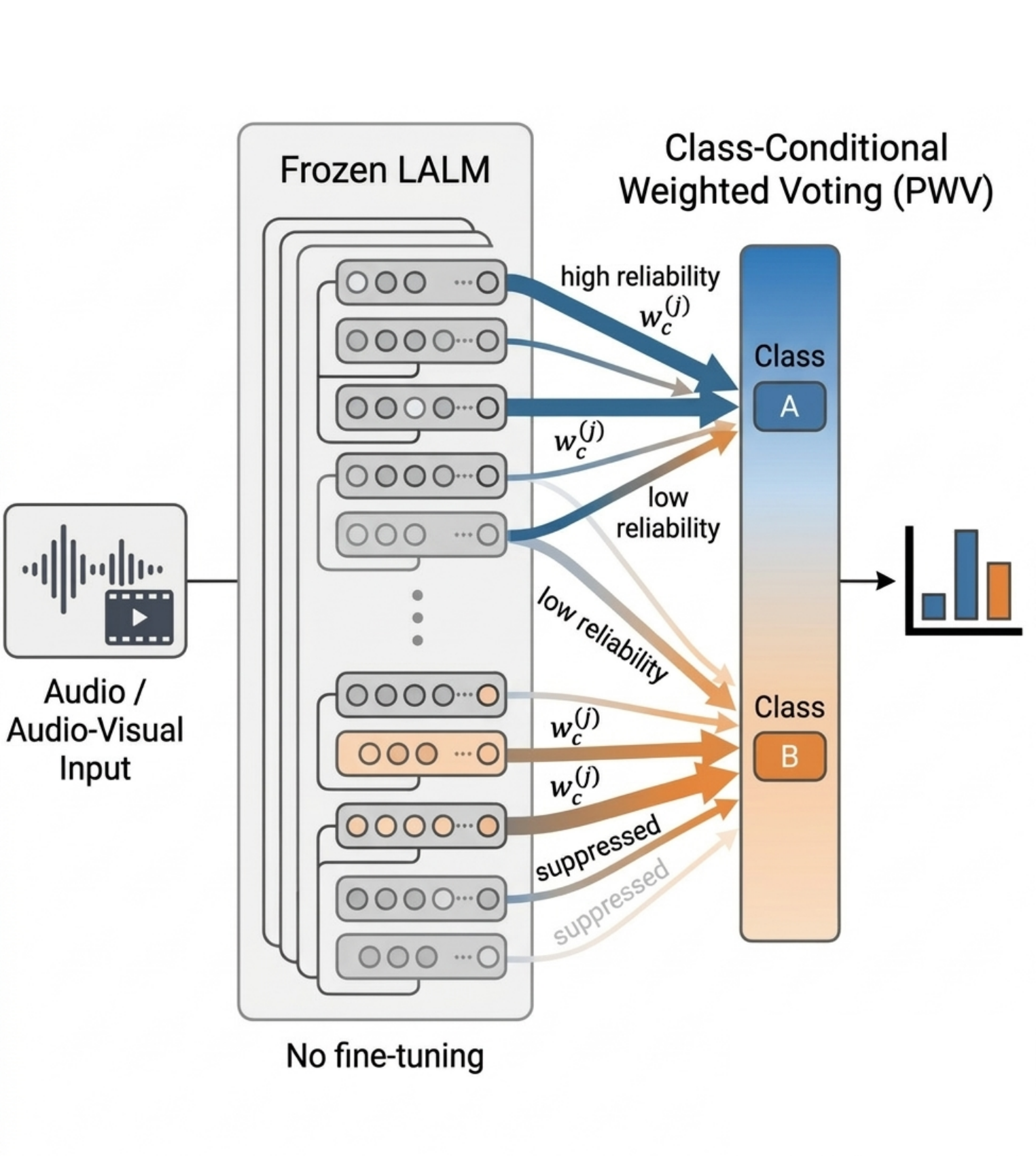}
    \caption{\textbf{Fine-tuning-free audio classification with CALM}. Given a frozen audio-language model, we extract per-head last-token hidden-state representations and compute class centroids. Each attention head produces cosine-similarity scores to all class centroids. For every head-class pair, we estimate a reliability score based on the margin to the next most confident class, which is visualized by the arrow width and color in the figure. Highly reliable heads contribute strongly (thick, saturated arrows), while unreliable heads are suppressed by lower weights. At inference time, class predictions are obtained via reliability-weighted soft voting across all heads. This yields accurate classification without task-specific fine-tuning.}
    \label{fig:teaser}
\end{figure}

\section{Introduction}
Large audio-language models (LALMs), such as Qwen2-Audio~\cite{chu2024qwen2audiotechnicalreport} and Qwen2.5-Omni~\cite{xu2025qwen25omnitechnicalreport}, have shown strong zero-shot and few-shot performance across a wide range of audio understanding tasks \cite{chu2024qwen2audiotechnicalreport,xu2025qwen25omnitechnicalreport}. Trained on large-scale paired audio-text data, these models generalize well without task-specific fine-tuning \cite{radford2022robustspeechrecognitionlargescale, huang2023audiogptunderstandinggeneratingspeech, lyu2023macawllmmultimodallanguagemodeling}.

Despite these strengths, repurposing LALMs for discriminative audio classification remains challenging. Specialized audio classifiers often outperform prompt-based approaches, but they require substantial labeled data and task-specific training pipelines \cite{gao2022wavpromptfewshotspokenlanguage, xie2019zeroshotaudioclassificationbased, elizalde2022claplearningaudioconcepts, guzhov2021audioclipextendingclipimage}. In contrast, directly leveraging frozen LALMs for classification offers reduced training cost \cite{hanif2024palmfewshotpromptlearning}, lower sample complexity in few-shot settings, and a unified model that supports both generative and discriminative use without retraining \cite{yang2024uniaudio15largelanguage}.

Recent work has shown that frozen generative models can be turned into strong discriminative classifiers by extracting intermediate representations rather than relying on prompt-based predictions~\cite{mitra2025enhancingfewshotvisionlanguageclassification, huang2024multimodaltaskvectorsenable, zhang2024visuallygroundedlanguagemodelsbad, hendel2023incontextlearningcreatestask, hojel2024findingvisualtaskvectors, todd2024functionvectorslargelanguage}. In particular, Sparse Attention Vectors (SAV)~\cite{mitra2025enhancingfewshotvisionlanguageclassification} show that a small subset of attention heads in a frozen audio-language model can serve as effective feature extractors for few-shot classification. The method selects the most discriminative heads and aggregates their predictions through majority voting, substantially narrowing the gap between generative models and specialized discriminative systems without requiring fine-tuning on labeled data.

In this paper, we extend this line of work to audio, audiovisual, and visual classification. Compared with vision or text, audio classification presents unique challenges, including multiple overlapping acoustic events \cite{dogan2024multilabelzeroshotaudioclassification, sridhar2024enhancingtemporalunderstandingaudio}, higher intra-class and inter-class variability, and temporal ambiguity in semantic cues. These characteristics make it unlikely that a single uniform aggregation strategy over attention heads is optimal across all classes.
While effective, SAV relies on a strong simplifying assumption: all selected attention heads are treated as \emph{equally reliable} across all classes. By contrast, prior work on interpretability and representation learning shows that attention heads exhibit \emph{functional specialization}, often responding differently to different semantic patterns or sensory stimuli~\cite{basile2025headpursuitprobingattention, jo-myaeng-2020-roles, ma2025cognitivemirrorsexploringdiverse, wang2024differentiationspecializationattentionheads}. This exposes a key limitation of uniform voting: weak or noisy heads can still influence predictions for classes where they are not discriminative \cite{qian2025headinformationbottleneck}.
To address this limitation, we propose \textbf{C}lass-Conditional Sparse \textbf{A}ttention Vectors for
Large Audio-\textbf{L}anguage \textbf{M}odels (\calm/), a principled extension of SAV that models attention heads as \emph{class-conditional experts} rather than uniformly reliable voters. \calm/ estimates a class-specific reliability score for each selected head from few-shot data and uses these scores in a reliability-weighted voting scheme at inference time.
Figure~\ref{fig:teaser} shows the overall pipeline.
Because \calm/ operates directly on a frozen LALM, it introduces only minimal computational overhead.
Our contributions are threefold:
\begin{enumerate}
    \item We introduce \calm/, a probabilistically weighted framework that models attention heads as class-conditional experts.
	    \item We present the first application of attention vectors as discriminative classifiers to audio, audiovisual, and visual classification, despite modality-specific challenges.
	    \item We demonstrate improvements over uniform-voting SAV baselines across audio, audiovisual, visual, and spoofing benchmarks, including ESC-50, VGGSound, AudioSet, EuroSAT, Oxford-IIIT Pets, and LA-Spoof.
\end{enumerate}


\section{Related Work}
\label{sec:previous_work}
\paragraph{Generative Models for Audio Classification.}
Recent work adapts large generative audio or multimodal models to classification through prompting, in-context demonstrations, or label-likelihood scoring \cite{taylor2025improvingaudioclassificationtransitioning,gong2024listenthinkunderstand,brown2020languagemodelsfewshotlearners,chen2019,kumar2023genzgenerativezeroshottext,hendrycks2021measuringmassivemultitasklanguage}. While appealing because they reuse a single frozen model, these approaches are often sensitive to prompt design and can lag behind specialized discriminative systems or adapted audio-language models \cite{olvera2024sounddescriptionexploringprompt,sahoo2025systematicsurveypromptengineering,salinas2024butterflyeffectalteringprompts,Choudhary_2022,deshmukh2024domainadaptationcontrastiveaudiolanguage}. Other methods improve classification through instruction tuning, preference optimization, or test-time adaptation, but they require additional optimization or supervision and therefore move away from the fine-tuning-free setting we target \cite{choi2025exploringfinetuninglargeaudio,bucher2024finetunedsmallllmsstill,ouyang2022traininglanguagemodelsfollow,sun19ttt,zhang2025aqattrlselfadaptationaudioquestion}.


\begin{figure*}[t!]
  \centering
  \includegraphics[width=0.85\textwidth]{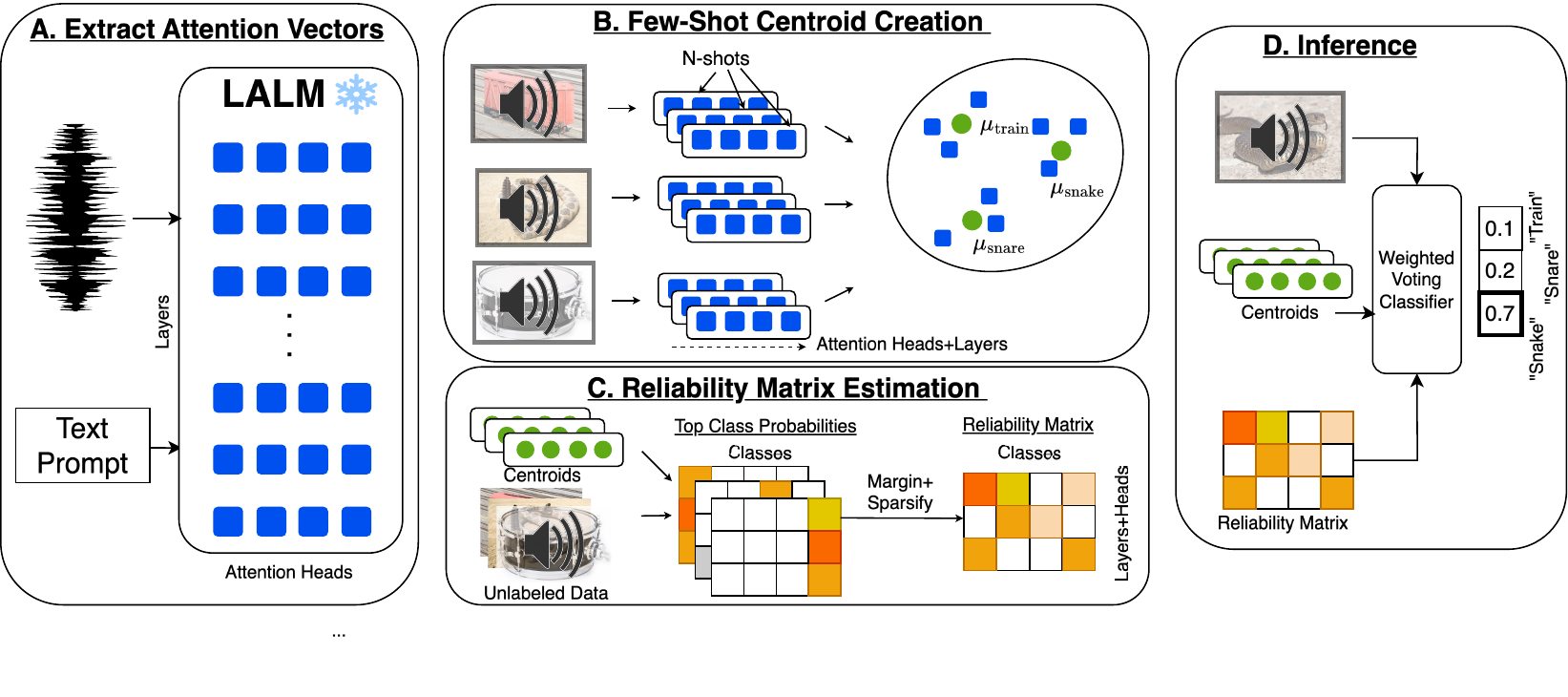}
      \vspace{-1.5em}

  \caption{\textbf{Overview of \calm/}: We extract attention-head vectors from a frozen LALM given an audio input and prompt. Then we build class centroids from few-shot training examples and estimate a class-conditional head reliability matrix using margin-based confidence and sparsifying to $k$ head experts per class. At inference, \calm/ combines centroid similarities with these reliability weights to perform weighted voting over heads. This allows for fine-tuning-free classification.}
    \label{fig:main-method}
\end{figure*}
\paragraph{Internal Representation of LALMs.}
A complementary direction uses frozen language or multimodal models as feature extractors rather than relying on decoded outputs \cite{belinkov2021probingclassifierspromisesshortcomings,hendel2023incontextlearningcreatestask,huang2024multimodaltaskvectorsenable,taylor2025improvingaudioclassificationtransitioning}. In audio, this includes representation-based classifiers formed from audio-text similarity in a shared embedding space, as well as methods that use intermediate hidden states for downstream prediction. Interpretability studies further show that attention heads exhibit functional specialization, suggesting that different heads capture different semantic factors \cite{clark2019doesbertlookat,voita2019analyzingmultiheadselfattentionspecialized,basile2025headpursuitprobingattention,jo-myaeng-2020-roles}. SAV \cite{mitra2025enhancingfewshotvisionlanguageclassification} turns this observation into a classifier by selecting a sparse set of discriminative heads and aggregating them with uniform voting. Our work builds directly on SAV, but replaces uniform aggregation with class-conditional reliability weighting so that different classes can rely on different expert heads.

\section{Methods}
\label{sec:methods}
We consider a $C$-way classification problem with a few-shot training set $\mathcal{D}_{\text{train}}=\{(x_i,c_i)\}_{i=1}^{N}$ and a test set $\mathcal{D}_{\text{test}}=\{(x_i,c_i)\}_{i=1}^{N_{\text{test}}}$, where $c_i\in\mathcal{C}=\{1,\dots,C\}$ and $n_c$ denotes the number of training examples for class $c$. Given an input $x$, we prompt a frozen audio-language or multimodal language model and extract the last-token representation from each attention head $j\in\{1,\dots,K\}$, denoted $h_j(x)\in\mathbb{R}^d$. We then build a fine-tuning-free classifier by computing class centroids in each head space and aggregating head-level evidence. Figure~\ref{fig:main-method} provides an overview.

We first briefly recap SAV \cite{mitra2025enhancingfewshotvisionlanguageclassification}, then introduce \calm/, our class-conditional weighting scheme. We finally describe a simple pseudo-labeling extension for unlabeled data.

\subsection{Recap: Sparse Attention Vectors}\label{subsec:sav}
The SAV classifier \cite{mitra2025enhancingfewshotvisionlanguageclassification} forms a nearest-centroid classifier in each attention head and aggregates the top-$k$ heads with uniform voting.
\paragraph{Per-head Centroids.}
For each class $c$ and head $j$, SAV computes the class centroid
\begin{equation}
\mu_c^{(j)} = \frac{1}{n_c}\sum_{i:c_i=c} h_j(x_i).
\label{centroid_avg}
\end{equation}
It scores a query $x$ against class $c$ in head $j$ using cosine similarity
\begin{equation}
s_j(x,c)
=
\frac{h_j(x)^\top \mu_c^{(j)}}{\lVert h_j(x)\rVert_2 \, \lVert \mu_c^{(j)}\rVert_2}.
\label{cos_sim}
\end{equation}
\paragraph{Top-$k$ head selection.}
Each head is ranked by its nearest-centroid classification accuracy on $\mathcal{D}_{\text{train}}$. Concretely, head $j$ predicts
\begin{equation}
\hat{y}^{(j)}(x) = \arg\max_{c\in\mathcal{C}}\;\;s_j(x,c),
\label{scoring_sav}
\end{equation}
and the top-$k$ heads define $\mathcal{H}_{\text{SAV}}$.

\paragraph{Majority Voting (Uniform Aggregation).}
At inference time, SAVs aggregates the selected heads by uniform voting:
\begin{equation}
\hat{y}(x_{\text{test}}) = \arg\max_{c\in\mathcal{C}} \sum_{j\in\mathcal{H}_{\text{SAV}}}\mathbf{1}\left[\hat{y}^{(j)}(x_{\text{test}})=c\right].
\label{uniform_voting}
\end{equation}
This uniform voting scheme assumes that all selected heads are equally reliable across all classes. \calm/ removes this assumption.

\subsection{\calm/}\label{subsec:calm}
\calm/ keeps the same per-head centroids as SAV, but replaces hard uniform voting with weighted aggregation of per-head posteriors. The key idea is to treat attention heads as \emph{class-specific experts}, so that classes can rely on different subsets of heads with different strengths.

\paragraph{Head Posteriors.} Using the per-head class centroids from Eq.~\ref{centroid_avg} and the similarity scores from Eq.~\ref{cos_sim}, we convert each head's score into a probability distribution over classes.
\begin{equation}
p_c^{(j)}(x) =
\frac{\exp\left(s_j(x,c)/\tau_p\right)}
{\sum_{c'\in\mathcal{C}}\exp\left(s_j(x,c')/\tau_p\right)},
\label{head_posterior}
\end{equation}
for some hyperparameter $\tau_p>0.$

\subsubsection{\calm/-Global: Head Weighting} \label{subsubsec:calm-global}
We first describe a simpler global-weighting variant, which assigns each head a single reliability shared across all classes. For a labeled example $(x_i,c_i)$, we define the clamped margin

\begin{equation}
m^{(j)}(x,c) = \max\Bigl(0, p_c^{(j)}(x) - \max_{c'\neq c}\;\; p_{c'}^{(j)}(x)\Bigr).
\label{margin}
\end{equation}

The clamp ensures that incorrect or low-confidence cases contribute zero. We then average margins over all training examples to obtain a single global reliability per head:
\begin{equation}
r^{(j)}=\frac{1}{N}\sum_{i=1}^{N} m^{(j)}(x_i,c_i).
\label{reliability_global}
\end{equation}
Retaining the top-$k$ heads with the largest reliabilities gives $\mathcal{H}_G$, and we normalize their weights as
\begin{equation}
w^{(j)} =
\begin{cases}
\displaystyle
\frac{\exp(r^{(j)} / \tau_w)}{\sum_{j' \in \mathcal{H}_G} \exp(r^{(j')} / \tau_w)}, & j \in \mathcal{H}_G, \\[1.0em]
0, & \text{otherwise.}
\end{cases}
\label{weight_global}
\end{equation}

At test time, we aggregate per-head posteriors using these global weights:
\begin{align}
p_c(x_{\text{test}})&=\sum_{j=1}^{K} w^{(j)}p_c^{(j)}(x_{\text{test}}),\label{inference_global}
\\
\hat{y}(x_{\text{test}})&=\arg\max_c\;\; p_c(x_{\text{test}}).
\label{inference_max}
\end{align}
This variant is useful for comparison, but it does not capture class-specific head specialization.

\subsubsection{Class-Conditional Head Weighting} \label{subsubsec:pwv-local}
Our full \calm/ model assigns a separate reliability to each head-class pair. After computing the margins in Eq.~\eqref{margin}, we average them class-wise:
\begin{equation}
r_c^{(j)}=\frac{1}{n_c}\sum_{i; c_i=c} m^{(j)}(x_i,c).
\label{reliability_pwv}
\end{equation}

Let the top-$k$ heads for class $c$ be $\mathcal{H}_c$. We normalize the class-conditional reliabilities with a softmax:

\begin{equation}
w_c^{(j)} =
\begin{cases}
\frac{\exp(r_c^{(j)} / \tau_w)}{\sum_{j' \in \mathcal{H}_c} \exp(r_c^{(j')} / \tau_w)}, & j \in \mathcal{H}_c, \\
0, & \text{otherwise,}
\end{cases}
\label{sparsity_pwv}
\end{equation}
for some hyperparameter $\tau_w>0.$ This generalizes the shared top-$k$ set and uniform voting used by SAV to a class-specific set of head experts.
\begin{equation}
p_c(x_{\text{test}})=\sum_{j=1}^{K} w_c^{(j)}p_c^{(j)}(x_{\text{test}}),
\label{eq}
\end{equation}
and predict with the same argmax rule as in Eq.~\eqref{inference_max}. Algorithm~\ref{alg:pwv} summarizes the class-conditional \calm/ procedure.

\begin{algorithm}[t]
\caption{CALM (class-conditional weighting)}
\label{alg:pwv}
\begin{algorithmic}[1]
\REQUIRE Frozen LALM, train set $\mathcal{D}=\{(x_i,y_i)\}$, num heads $K$, and sparsity $k$
\STATE Extract attention-head features $h_j(x_i)$ for all $j \in \{1,\dots,K\}$
\STATE Compute class centroids $\mu_c^{(j)} = \frac{1}{n_c}\sum_{i:y_i=c} h_j(x_i)$
\FOR{each class $c$ and head $j$}
    \STATE Compute posteriors $p_c^{(j)}(x_i)$ via Eq.~\ref{head_posterior}
    \STATE Compute margins via Eq.~\ref{margin}
    \STATE $r_c^{(j)} \leftarrow \frac{1}{n_c}\sum_{i:y_i=c} m_i^{(j)}$
\ENDFOR
\STATE Select top-$k$ reliable heads per class and normalize weights $w_c^{(j)}$
\STATE Predict $\hat{y}=\arg\max_c \sum_j w_c^{(j)} p_c^{(j)}(x_{\text{test}})$
\end{algorithmic}
\end{algorithm}

\subsection{Self-supervised \calm/}
\label{sec:semi_supervised}
To use \calm/ without ground-truth labels, we generate $M$ stochastic predictions for each unlabeled example using decoding temperatures $\tau \sim \mathcal{U}(1, 2.5)$. We assign a pseudo-label by majority vote, retain only examples where at least $50\%$ of rollouts agree, and then run SAV or \calm/ on the retained pseudo-labeled set $\tilde{\mathcal{D}}_{\text{train}}=\{(x_i,\tilde{c}_i)\}$.

\begin{table*}[t!]
\centering
\footnotesize
\setlength{\tabcolsep}{4pt}
\renewcommand{\arraystretch}{1.02}
\begin{tabular*}{0.96\textwidth}{@{\extracolsep{\fill}}lccccc@{}}
\toprule
\textbf{Dataset} & \textbf{ZS} & \textbf{SAV} & \textbf{SAV+PL} & \textbf{\calm/} & \textbf{\calm/+PL} \\
\midrule
\multicolumn{6}{@{}l}{\textbf{Qwen2-Audio}} \\
\midrule
ESC-50   & 73.75 & 99.00 & 96.50 & \textbf{99.00} & 98.75 \\
VGGSound & 72.94 & 85.64 & 85.65 & \textbf{88.15} & 87.67 \\
AudioSet & 46.59 & 44.88 & 49.06 & \textbf{59.40} & 51.80 \\
LA-Spoof & 16.19 & \textbf{72.00} & 70.03 & 61.10 & 22.73 \\
\midrule
\multicolumn{6}{@{}l}{\textbf{Qwen2.5-Omni}} \\
\midrule
ESC-50   & 86.00 & 99.00 & 91.00 & \textbf{99.25} & 98.50 \\
VGGSound & 61.57 & 87.11 & 81.78 & \textbf{87.85} & 86.28 \\
AudioSet & 61.52 & 69.52 & 63.54 & \textbf{70.50} & 66.70 \\
VGGVideo & 70.71 & 89.49 & --    & \textbf{91.02} & --    \\
LA-Spoof & 15.82 & 73.64 & 67.01 & \textbf{81.99} & 70.13 \\
EuroSAT  & 59.67 & 68.72 & 58.10 & \textbf{72.80} & 62.70 \\
Pets     & 96.05 & 98.27 & 97.92 & \textbf{99.23} & 97.16 \\
\bottomrule
\end{tabular*}
\caption{\textbf{Main classification results.} Accuracy (\%) for zero-shot (ZS), SAV, and \calm/ with and without pseudo-labeling (PL). Best results within each model block are shown in \textbf{bold}.}
\label{table:main_results}
\end{table*}

\section{Experiments and Results}
\label{sec:experiments}

We evaluate our method on a range of audio and audiovisual classification benchmarks. We first describe the implementation details, models, datasets, and evaluation protocols, and then analyze the impact of head sparsity and class-specific weighting through ablations.

\subsection{Implementation Details}
All experiments are implemented in PyTorch and run on NVIDIA A6000 GPUs. We use frozen pretrained models from the Qwen family~\cite{chu2024qwen2audiotechnicalreport,xu2025qwen25omnitechnicalreport}.
\emph{Qwen2-Audio}~\cite{chu2024qwen2audiotechnicalreport} is an audio-language model in which an audio encoder feeds into the language model. The language model consists of 32 Transformer layers with 32 attention heads per layer (1024 heads total), each with a hidden dimension of 128. We extract attention head activations from the language model layers rather than the audio encoder to capture cross-modal audio-text interactions.
\emph{Qwen2.5-Omni}~\cite{xu2025qwen25omnitechnicalreport} is a multimodal model supporting audio, image, video, and speech inputs. 
As with Qwen2-Audio, we extract attention activations exclusively from the language tower.
To select $\tau_p$ and $\tau_w$, we perform a hyperparameter sweep either on a small labeled validation split or purely from pseudo-labeled data (see \hyperref[par:hp_search_unlabeled]{Hyperparameter Selection without Labels}). We sweep $\tau_p, \tau_w \in \{0.001, 0.03, 0.1, 0.5, 1, 2\}$. We use $\tau_p = 0.03$ and $\tau_w = 0.5$ for almost all datasets; for AudioSet, we use $\tau_p = 0.001$ and $\tau_w = 0.5$. For ESC-50, VGGSound, and AudioSet, an example prompt is: \texttt{What caption does the given audio belong to? A. washing\_machine B. toilet\_flush C. car\_horn D. wind}. For spoofing detection, an example prompt is: \texttt{Is this audio bona fide or spoofed? A. bona\_fide B. spoofed}. We use the same input-formatting logic for image inputs.

\subsection{Datasets}
We primarily evaluate our approach on audio classification benchmarks, and further extend it to audiovisual recognition and audio spoofing detection. Across all datasets, we follow the input formatting protocol of SAV~\cite{mitra2025enhancingfewshotvisionlanguageclassification}, casting each task into a multiple-choice question-answering formulation suitable for LALMs.


\noindent\textbf{ESC-50}~\cite{piczak2015dataset} is a standard environmental sound classification benchmark consisting of 2,000 five-second audio clips across 50 classes. We sample class-balanced subsets according to the desired shot setting and frame the task as multiple-choice audio classification.

\noindent\textbf{VGGSound}~\cite{Chen20} is a large-scale audiovisual dataset containing over 200k videos spanning 309 sound event classes. Compared to ESC-50, it exhibits substantially greater semantic and acoustic diversity. We adopt the same multiple-choice formulation and use both audio and video modalities for audiovisual experiments.

\noindent\textbf{AudioSet}~\cite{7952261} is a large-scale ontology-based audiovisual dataset with 527 non-mutually exclusive sound event classes. Due to its multi-label nature, we decompose each audio clip into independent classification queries, one per ground-truth label, and report strict \emph{exact-match} accuracy at the clip level.

\noindent\textbf{ASVspoof}~\cite{wang2020asvspoof2019largescalepublic} is used for spoofing detection. We study the LA (Logical Access) subset of ASVspoof 2019, which requires distinguishing bona fide speech from synthetic audio generated via speech synthesis and voice conversion methods. Given the class imbalance, we report Macro-F1 as the primary evaluation metric~\cite{zhang2025multiapispoofmultiapidataset}.

\noindent\textbf{EuroSAT}~\cite{helber2019eurosatnoveldatasetdeep} is a vision classification dataset for land-cover recognition. It contains 10 classes with approximately 2,000--3,000 images per class. We randomly sample a 20-shot support set and use the remaining images for evaluation.

\noindent\textbf{Oxford-IIIT Pets}~\cite{parkhi12a} is another vision classification dataset, containing 37 cat and dog breeds. The main challenge is the subtle visual differences between breeds. We follow the same multiple-choice protocol as above, pairing the ground-truth breed with three randomly sampled distractor breeds.

\subsection{Results}

\label{sec:results}
\subsubsection{Main Results}

Table~\ref{table:main_results} summarizes the main comparison against zero-shot inference, SAV, and the pseudo-labeling variant from Section~\ref{sec:semi_supervised}. Across both LALMs and most datasets, \calm/ improves over uniform SAV aggregation, with the largest gains appearing on the most challenging tasks. On ESC-50, both methods are near saturation, so the margin is small. On VGGSound, \calm/ improves over SAV by +2.51 points on Qwen2-Audio and +0.74 points on Qwen2.5-Omni. The strongest gain appears on AudioSet, where \calm/ improves over SAV by +14.52 points on Qwen2-Audio, suggesting that class-conditional weighting is especially useful when the label space is large and heterogeneous.

The same trend extends beyond audio-only classification. On Qwen2.5-Omni, \calm/ improves over SAV on VGGVideo, EuroSAT, and Oxford-IIIT Pets, showing that class-conditional head weighting transfers not only to audiovisual recognition but also to pure image classification. LA-Spoof is the main exception: \calm/ improves over SAV for Qwen2.5-Omni but underperforms for Qwen2-Audio. This is consistent with the binary nature of spoofing detection, where estimating separate class-specific weights from few examples can overfit more easily than in large multi-class problems.

Pseudo-labeling is best viewed as a complementary self-training setting rather than the main source of \calm/'s gains. As shown in Table~\ref{table:main_results}, \calm/+PL is usually below fully supervised \calm/, although it remains competitive on most settings where additional unlabeled data helps stabilize the centroids.

\paragraph{Comparison to Fine-tuning.}
Table~\ref{tab:lora_comparison} compares \calm/ against standard fine-tuning techniques such as linear probing and LoRA, and additional few-shot baselines such as nearest neighbors using CLAP embeddings~\cite{elizalde2022claplearningaudioconcepts}. \calm/ matches the best result on ESC-50, improves over both LoRA and SAV on VGGSound, and remains substantially stronger than CLAP. The main failure case is again LA-Spoof, where discriminative fine-tuning is more effective. 
\begin{table}[t]
\centering
\small
\setlength{\tabcolsep}{3.2pt}
\renewcommand{\arraystretch}{1.07}
\begin{tabular}{@{}lccc@{}}
\toprule
\textbf{Method} & \textbf{ESC-50} & \textbf{VGGSound} & \textbf{LA-Spoof F1} \\
\midrule
CLAP & 95.50 & 35.12 & 59.79 \\
Linear Probe & 97.50 & 81.48 & \textbf{85.13} \\
LoRA & 98.00 & 87.23 & 69.68 \\
SAV & 99.00 & 85.64 & 72.00 \\
CALM & \textbf{99.00} & \textbf{88.15} & 61.10 \\
\bottomrule
\end{tabular}
\caption{\textbf{Comparison to stronger baselines.} CALM remains competitive with linear probing and LoRA while outperforming SAV on VGGSound.}
\label{tab:lora_comparison}
\end{table}

\subsubsection{Class-Specific Head Specialization}\label{sec:head_specialization}
\begin{figure*}[t!]
  \centering
  \begin{minipage}[t]{0.48\textwidth}
    \centering
    \includegraphics[width=0.94\linewidth]{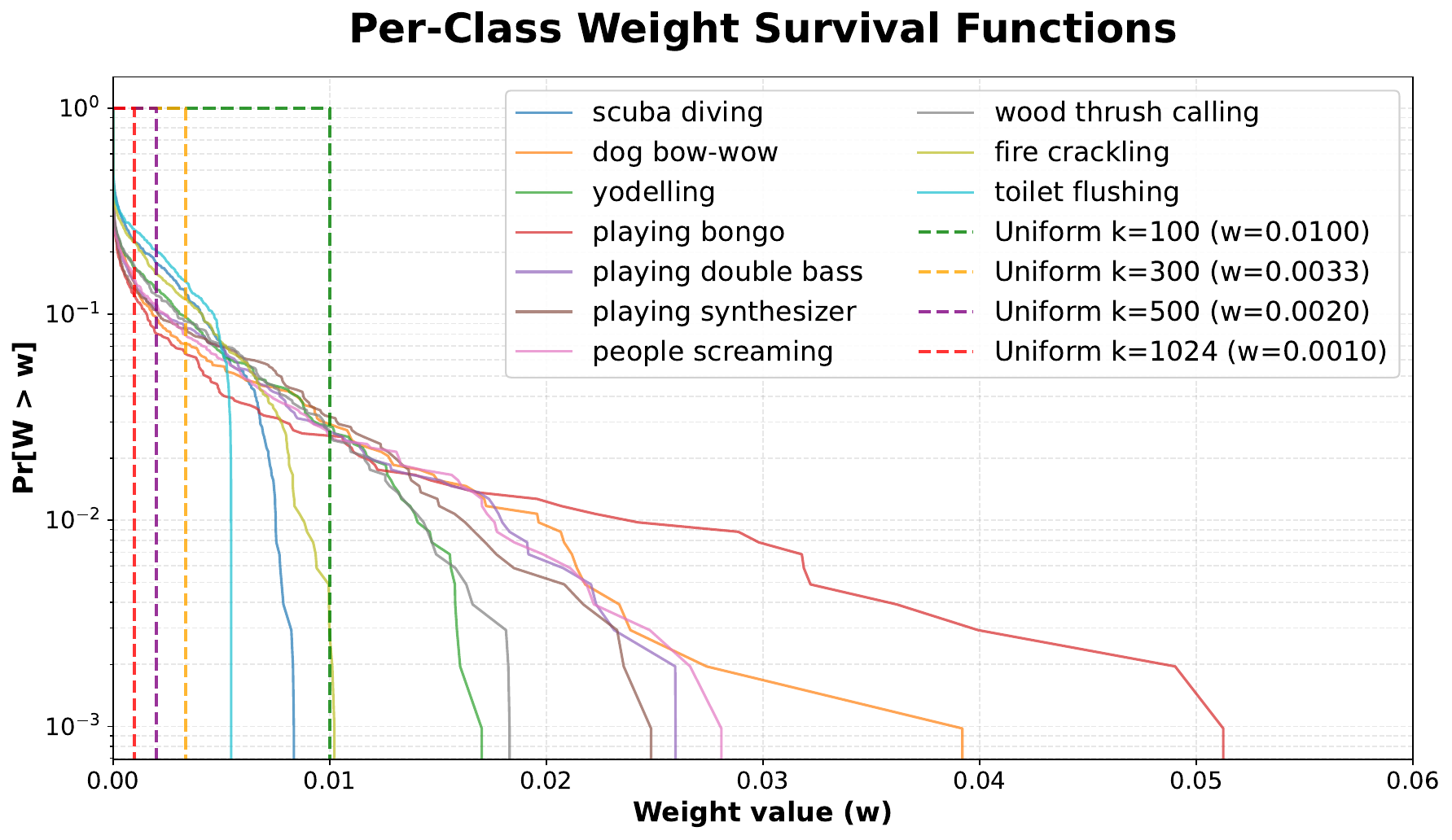}
    \captionof{figure}{\textbf{Per-class weight survival functions for VGGSound.} Different classes concentrate weight on different subsets of attention heads. Dashed lines show the corresponding uniform-voting baselines for $k = 100, 300, 500,$ and $1024$ heads.}
    \label{fig:weight_survival}
  \end{minipage}
  \hfill
  \begin{minipage}[t]{0.48\textwidth}
    \centering
    \includegraphics[width=0.94\linewidth]{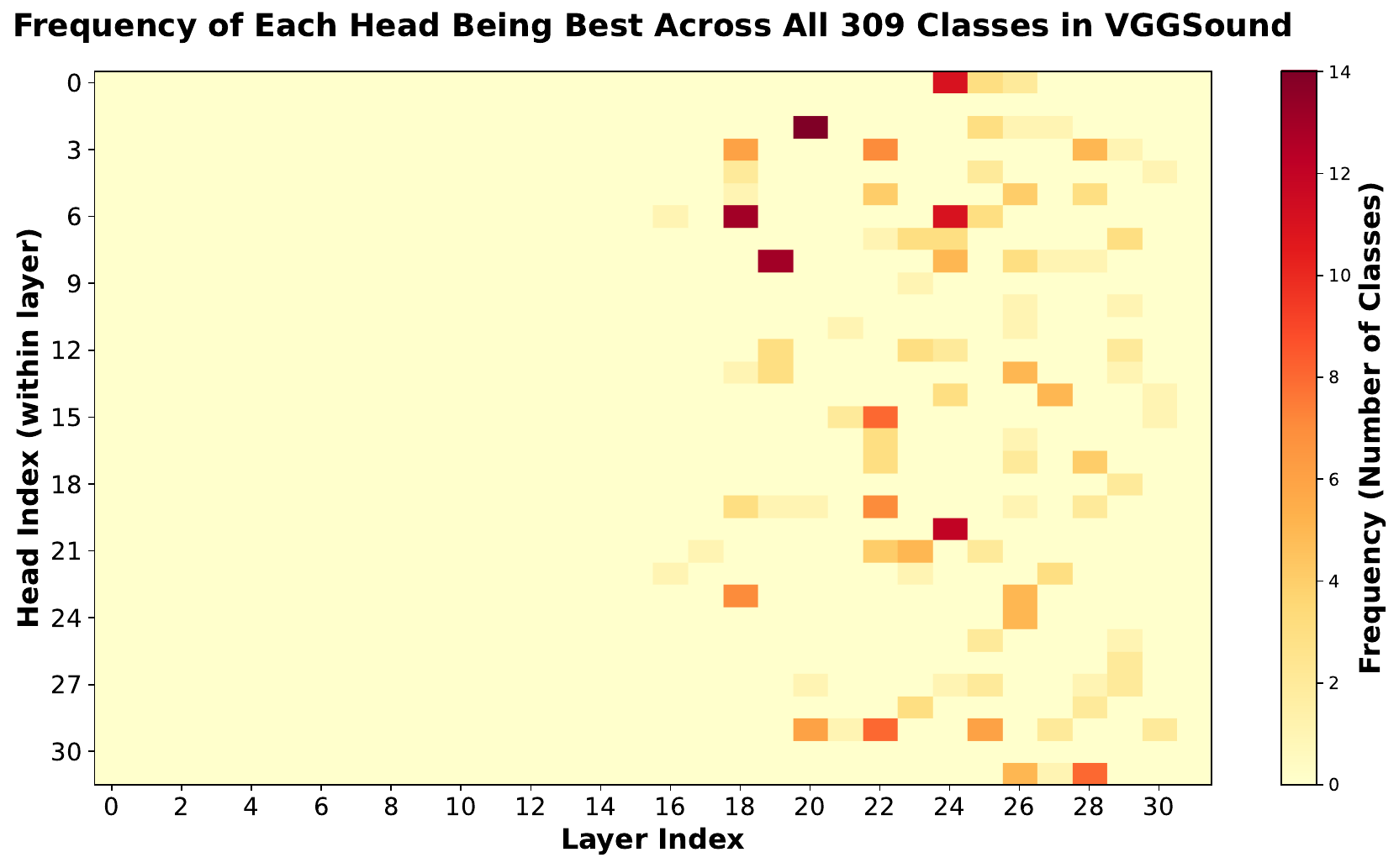}
    \captionof{figure}{\textbf{Most influential attention heads across classes.} Heatmap of the most influential head (highest $w^{(j)}$) for each VGGSound class.}
    \label{fig:expert-heads}
  \end{minipage}
  \figvsp
\end{figure*}

To better understand head specialization, Figure~\ref{fig:weight_survival} visualizes per-class weight survival functions for representative VGGSound classes, while Figure~\ref{fig:expert-heads} shows the most influential head for each class across the 309 VGGSound classes. Under SAV, all selected heads receive the same weight $\frac{1}{k}$, shown by the dashed baselines in Figure~\ref{fig:weight_survival}. In contrast, \calm/ learns class-conditional, non-uniform weightings: the survival curves differ substantially across classes, and several have heavy tails, indicating that a small subset of heads carries most of the evidence. Figure~\ref{fig:expert-heads} further shows that the strongest heads concentrate in later layers, suggesting that LALMs contain both class-specific experts and a smaller set of broadly useful heads and that CALM exploits structured specialization rather than diffuse contributions from all heads. 

\begin{figure*}[t!]
  \centering
  \begin{subfigure}[t]{0.48\linewidth}
    \centering
    \includegraphics[width=0.94\linewidth]{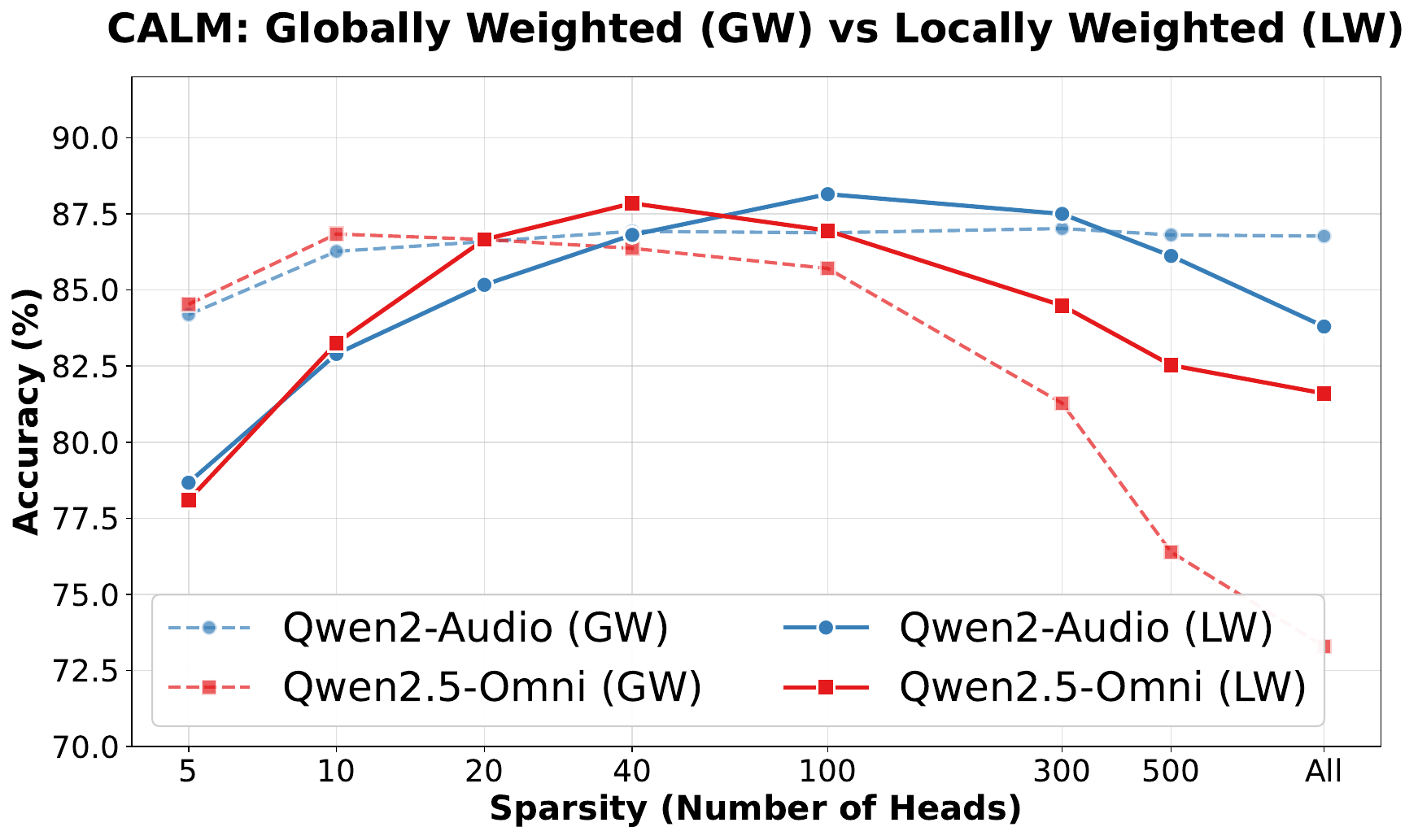}
  \end{subfigure}
  \hfill
  \begin{subfigure}[t]{0.48\linewidth}
    \centering
    \includegraphics[width=0.94\linewidth]{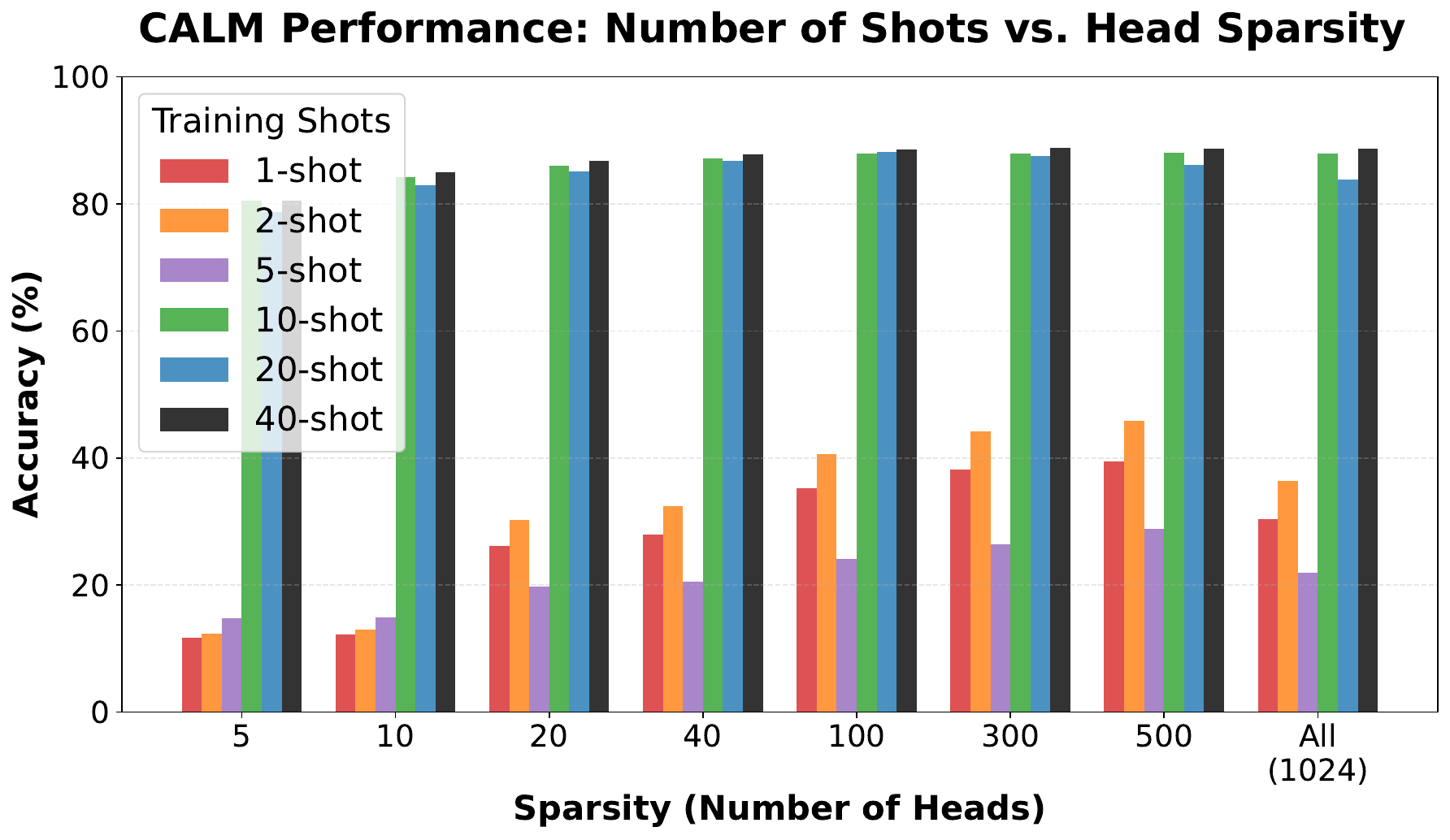}
  \end{subfigure}
  \caption{\textbf{CALM ablations.} \textbf{Left:} class-conditional local weighting consistently outperforms global weighting, especially when more heads are retained, supporting the hypothesis that attention heads specialize by class. \textbf{Right:} performance improves with more training shots and typically peaks at intermediate sparsity levels, showing that CALM benefits from both better centroid estimates and selective head pruning.}
  \label{fig:pwv_ablations}
  \figvsp
\end{figure*}

\subsubsection{Ablations}

We next validate the two main design choices in \calm/: class-conditional weighting and sparse reliability-based aggregation.

\paragraph{Class-Conditional vs Global Weighting.}
Table~\ref{tab:pwv_algorithmic_ablation} and the left panel of Figure~\ref{fig:pwv_ablations} compare the full class-conditional method with the simpler global variant from Section~\ref{subsubsec:calm-global}. Local weighting is consistently better on VGGSound, where class-specific head specialization matters most, while global weighting is better on LA-Spoof. This reinforces the interpretation from Table~\ref{table:main_results}: in a binary setting with limited support data, sharing one reliability score per head can regularize better than fitting separate class-wise weights. The pattern is also consistent with the head specialization analysis in Section~\ref{sec:head_specialization}: if different classes repeatedly route through different late-layer heads, then class-conditional weighting should outperform any scheme that forces all classes to share a single reliability profile.
\paragraph{Component Ablations.}
Table~\ref{tab:pwv_component_ablation} isolates the effect of the two main ingredients in \calm/. Removing L2 normalization consistently hurts performance, which suggests that raw head activations have scale variation that degrades centroid quality. Removing margin-based reliability also lowers accuracy, especially for Qwen2.5-Omni, showing that selecting heads by discriminability is important in addition to sparse aggregation. Together, these results show that both normalization and reliability weighting are necessary parts of the final method.

\begin{table}[t]
\centering
\small
\setlength{\tabcolsep}{4pt}
\renewcommand{\arraystretch}{1.1}
\resizebox{0.8\columnwidth}{!}{%
\begin{tabular}{@{}lcc@{}}
\toprule
\multicolumn{3}{c}{\textbf{Algorithmic Ablations}} \\
\midrule
\textbf{Model} & \textbf{VGGSound} & \textbf{LA-Spoof} \\
\midrule
\multicolumn{3}{l}{\textit{Global}} \\
Qwen2-Audio    & 87.02 & \textbf{69.07} \\
Qwen2.5-Omni   & 86.84 & \textbf{83.60} \\
\midrule
\multicolumn{3}{l}{\textit{Class-cond.}} \\
Qwen2-Audio    & \textbf{88.15} & 61.10 \\
Qwen2.5-Omni   & \textbf{87.85} & 81.99 \\
\bottomrule
\end{tabular}
}

\caption{\textbf{Algorithmic ablations.} Comparison between global and class-conditional weighting on VGGSound and LA-Spoof.}
\label{tab:pwv_algorithmic_ablation}
\figvsp
\end{table}

\begin{table}[t!]
\centering
\small
\setlength{\tabcolsep}{6pt}
\renewcommand{\arraystretch}{1.15}
\resizebox{\columnwidth}{!}{%
\begin{tabular}{@{}lcc@{}}
\toprule
\multicolumn{3}{c}{\textbf{L2 Norm and Margin Weighting}} \\
\midrule
\textbf{Method} & \textbf{Qwen2-Audio} & \textbf{Qwen2.5-Omni} \\
\midrule
SAV Baseline         & 85.64          & 87.11          \\
\midrule
\calm/                 & \textbf{88.15} & \textbf{87.85} \\
-- L2 Norm          & 87.03          & 87.00          \\
-- Margin           & 87.33          & 85.52          \\
\bottomrule
\end{tabular}
}
\caption{\textbf{\calm/ component ablations.} Effect of removing L2 normalization or margin-based reliability weighting.}
\label{tab:pwv_component_ablation}
\figvsp
\end{table}

\paragraph{Shot and Sparsity Sensitivity.}
The right panel of Figure~\ref{fig:pwv_ablations} shows that CALM improves as the number of support examples increases, with the largest gains appearing between 5 and 10 shots. The same plot also shows that the best performance is usually obtained with intermediate sparsity rather than either extreme, which supports the central design intuition: too few heads discard useful evidence, while too many heads dilute the contribution of the most discriminative experts.

\paragraph{Random-head Counterfactual.}
Table~\ref{tab:random_head_counterfactual} tests whether CALM's gains come from many heads or from structured selection and weighting. Random head selection with random weights degrades performance sharply in sparse regimes (27.35\% vs. 78.10\% at top-$k=5$ on VGGSound), and CALM remains significantly better across all $k$.
\begin{table}[t]
\centering
\small
\renewcommand{\arraystretch}{1.14}
\begin{tabular*}{0.72\columnwidth}{@{\extracolsep{\fill}}lcc@{}}
\toprule
\textbf{top-$k$} & \textbf{Random} & \textbf{CALM} \\
\midrule
5   & 27.35 & \textbf{78.10} \\
10  & 38.04 & \textbf{83.26} \\
20  & 53.08 & \textbf{86.66} \\
40  & 68.52 & \textbf{87.85} \\
100 & 79.79 & \textbf{86.95} \\
\bottomrule
\end{tabular*}
\caption{\textbf{Random-head counterfactual.} Results with Qwen2.5-Omni on VGGSound.}
\label{tab:random_head_counterfactual}
\end{table}

\paragraph{Seed Stability.}

Table~\ref{tab:stability_seeds} measures robustness across 10 random 20-shot VGGSound splits on Qwen2-Audio. CALM consistently outperforms SAV across all tested top-$k$ values with low variance, showing stable gains across support-set sampling and sparsity choices.
\begin{table}[t]
\centering
\small
\renewcommand{\arraystretch}{1.14}
\begin{tabular*}{0.72\columnwidth}{@{\extracolsep{\fill}}lcc@{}}
\toprule
\textbf{top-$k$} & \textbf{SAV} & \textbf{CALM} \\
\midrule
20  & 85.22 $\pm$ 0.18 & \textbf{86.38 $\pm$ 0.16} \\
40  & 85.20 $\pm$ 0.24 & \textbf{87.46 $\pm$ 0.13} \\
100 & 85.37 $\pm$ 0.19 & \textbf{88.22 $\pm$ 0.10} \\
300 & 85.65 $\pm$ 0.12 & \textbf{88.45 $\pm$ 0.11} \\
\bottomrule
\end{tabular*}
\caption{\textbf{Seed stability.} Results with Qwen2-Audio on VGGSound across 10 random 20-shot splits.}
\label{tab:stability_seeds}
\end{table}

\begin{table}[t]
\centering
\small
\renewcommand{\arraystretch}{1.14}
\begin{tabular*}{0.72\columnwidth}{@{\extracolsep{\fill}}lcc@{}}
\toprule
\textbf{Backbone} & \textbf{SAV} & \textbf{CALM} \\
\midrule
Qwen2-VL-7B & 71.03 & \textbf{73.93} \\
Phi-4-Multimodal  & 47.99 & \textbf{49.72} \\
\bottomrule
\end{tabular*}
\caption{\textbf{Cross-architecture transfer}. Results with Qwen2-VL-7B and Phi-4-Multimodal on VGGSound.}
\label{tab:cross_arch_transfer}
\end{table}
\paragraph{Hyperparameter Selection without Labels.}\label{par:hp_search_unlabeled}
Hyperparameters can be selected using either a small labeled validation split or selecting hyper-parameters on zero-shot pseudo-labeled test-data. 
On VGGSound, pseudo-label-selected hyperparameters match labeled-validation selection and yield 88.15\% test accuracy, improving over the 72.94\% zero-shot baseline by +15.21 points without labeled tuning.

\paragraph{Cross-architecture Transfer.}
To test transfer beyond our main LALMs, we apply CALM to Qwen2-VL-7B and Phi-4-Multimodal~\cite{abouelenin2025phi4mini} (Table~\ref{tab:cross_arch_transfer}). CALM improves over SAV on both settings (71.03\%$\rightarrow$73.93\% and 47.99\%$\rightarrow$49.72\%), suggesting benefits across architectures and modality combinations.



\section{Conclusion}

We presented \textbf{CALM}, a fine-tuning-free classifier that treats attention heads in frozen LALMs as class-conditional experts. Across audio, audiovisual, and visual benchmarks, CALM improves over uniform SAV voting. Ablations show that class-specific margin-based reliability and normalization are key to these gains, indicating that reliability-aware sparse aggregation exposes discriminative structure already present in frozen multimodal models.

\section*{Limitations}

\calm/ remains sensitive to label scarcity and label noise because class-specific reliabilities are estimated from small support sets, reflected in weaker binary spoofing results and the gap between supervised and pseudo-labeled settings. The method also introduces dataset-dependent temperature hyperparameters $(\tau_p, \tau_w)$ and validation-dependent sparsity selection, though these can be chosen \emph{from unlabeled} data as in CALM+PL. Finally, we evaluate only a limited set of frozen backbones and benchmarks; broader architecture, modality, and task coverage is needed to fully assess generality.

\FloatBarrier
\bibliography{custom}

@string{nips = "NeurIPS"}

@string{icml = "ICML"}

@string{iclr = "ICLR"}

@string{acl = "ACL"}

@string{emnlp = "EMNLP"}

@string{iccv = "ICCV"}

@string{eccv = "ECCV"}

@string{interspeech = "Interspeech"}

@string{icassp = "ICASSP"}

@inproceedings{mitra2025enhancingfewshotvisionlanguageclassification,
      title={Enhancing Few-Shot Vision-Language Classification with Large Multimodal Model Features}, 
      author={Chancharik Mitra and Brandon Huang and Tianning Chai and Zhiqiu Lin and Assaf Arbelle and Rogerio Feris and Leonid Karlinsky and Trevor Darrell and Deva Ramanan and Roei Herzig},
      booktitle=iccv,
      year={2025},
      eprint={2412.00142},
      archivePrefix={arXiv},
      primaryClass={cs.CV},
      url={https://arxiv.org/abs/2412.00142}, 
}

@misc{chu2024qwen2audiotechnicalreport,
      title={Qwen2-Audio Technical Report}, 
      author={Yunfei Chu and Jin Xu and Qian Yang and Haojie Wei and Xipin Wei and Zhifang Guo and Yichong Leng and Yuanjun Lv and Jinzheng He and Junyang Lin and Chang Zhou and Jingren Zhou},
      year={2024},
      eprint={2407.10759},
      archivePrefix={arXiv},
      primaryClass={eess.AS},
      url={https://arxiv.org/abs/2407.10759}, 
}

@misc{xu2025qwen25omnitechnicalreport,
      title={Qwen2.5-Omni Technical Report}, 
      author={Jin Xu and Zhifang Guo and Jinzheng He and Hangrui Hu and Ting He and Shuai Bai and Keqin Chen and Jialin Wang and Yang Fan and Kai Dang and Bin Zhang and Xiong Wang and Yunfei Chu and Junyang Lin},
      year={2025},
      eprint={2503.20215},
      archivePrefix={arXiv},
      primaryClass={cs.CL},
      url={https://arxiv.org/abs/2503.20215}, 
}

@inproceedings{piczak2015dataset,
  title = {{ESC}: {Dataset} for {Environmental Sound Classification}},
  author = {Piczak, Karol J.},
  booktitle = {Proceedings of the 23rd {Annual ACM Conference} on {Multimedia}},
  year = {2015},
  date = {2015-10-13},
  url = {http://dl.acm.org/citation.cfm?doid=2733373.2806390},
  doi = {10.1145/2733373.2806390},
  location = {{Brisbane, Australia}},
  isbn = {978-1-4503-3459-4},
  publisher = {{ACM Press}},
  pages = {1015--1018}
}

@InProceedings{Chen20,
  author       = "Honglie Chen and Weidi Xie and Andrea Vedaldi and Andrew Zisserman",
  title        = "VGGSound: A Large-scale Audio-Visual Dataset",
  booktitle    = "International Conference on Acoustics, Speech, and Signal Processing (ICASSP)",
  year         = "2020",
}

@INPROCEEDINGS{7952261,
  author={Gemmeke, Jort F. and Ellis, Daniel P. W. and Freedman, Dylan and Jansen, Aren and Lawrence, Wade and Moore, R. Channing and Plakal, Manoj and Ritter, Marvin},
  booktitle={2017 IEEE International Conference on Acoustics, Speech and Signal Processing (ICASSP)}, 
  title={Audio Set: An ontology and human-labeled dataset for audio events}, 
  year={2017},
  volume={},
  number={},
  pages={776-780},
  doi={10.1109/ICASSP.2017.7952261}}

@article{wang2020asvspoof2019largescalepublic,
      title={ASVspoof 2019: A large-scale public database of synthesized, converted and replayed speech}, 
      author={Xin Wang and Junichi Yamagishi and Massimiliano Todisco and Hector Delgado and Andreas Nautsch and Nicholas Evans and Md Sahidullah and Ville Vestman and Tomi Kinnunen and Kong Aik Lee and Lauri Juvela and Paavo Alku and Yu-Huai Peng and Hsin-Te Hwang and Yu Tsao and Hsin-Min Wang and Sebastien Le Maguer and Markus Becker and Fergus Henderson and Rob Clark and Yu Zhang and Quan Wang and Ye Jia and Kai Onuma and Koji Mushika and Takashi Kaneda and Yuan Jiang and Li-Juan Liu and Yi-Chiao Wu and Wen-Chin Huang and Tomoki Toda and Kou Tanaka and Hirokazu Kameoka and Ingmar Steiner and Driss Matrouf and Jean-Francois Bonastre and Avashna Govender and Srikanth Ronanki and Jing-Xuan Zhang and Zhen-Hua Ling},
      journal={Computer Speech and Language},
      year={2020},
      url={https://arxiv.org/abs/1911.01601}, 
}

@inproceedings{radford2022robustspeechrecognitionlargescale,
      title={Robust Speech Recognition via Large-Scale Weak Supervision},
      booktitle=icml,
      author={Alec Radford and Jong Wook Kim and Tao Xu and Greg Brockman and Christine McLeavey and Ilya Sutskever},
      year={2022},
      url={https://arxiv.org/abs/2212.04356}, 
}

@misc{huang2023audiogptunderstandinggeneratingspeech,
      title={AudioGPT: Understanding and Generating Speech, Music, Sound, and Talking Head}, 
      author={Rongjie Huang and Mingze Li and Dongchao Yang and Jiatong Shi and Xuankai Chang and Zhenhui Ye and Yuning Wu and Zhiqing Hong and Jiawei Huang and Jinglin Liu and Yi Ren and Zhou Zhao and Shinji Watanabe},
      year={2023},
      eprint={2304.12995},
      archivePrefix={arXiv},
      primaryClass={cs.CL},
      url={https://arxiv.org/abs/2304.12995}, 
}

@misc{lyu2023macawllmmultimodallanguagemodeling,
      title={Macaw-LLM: Multi-Modal Language Modeling with Image, Audio, Video, and Text Integration}, 
      author={Chenyang Lyu and Minghao Wu and Longyue Wang and Xinting Huang and Bingshuai Liu and Zefeng Du and Shuming Shi and Zhaopeng Tu},
      year={2023},
      eprint={2306.09093},
      archivePrefix={arXiv},
      primaryClass={cs.CL},
      url={https://arxiv.org/abs/2306.09093}, 
}

@inproceedings{gao2022wavpromptfewshotspokenlanguage,
      title={WAVPROMPT: Towards Few-Shot Spoken Language Understanding with Frozen Language Models}, 
      author={Heting Gao and Junrui Ni and Kaizhi Qian and Yang Zhang and Shiyu Chang and Mark Hasegawa-Johnson},
      year={2022},
      booktitle=interspeech,
      url={https://arxiv.org/abs/2203.15863}, 
}

@misc{xie2019zeroshotaudioclassificationbased,
      title={Zero-Shot Audio Classification Based on Class Label Embeddings}, 
      author={Huang Xie and Tuomas Virtanen},
      year={2019},
      eprint={1905.01926},
      archivePrefix={arXiv},
      primaryClass={cs.LG},
      url={https://arxiv.org/abs/1905.01926}, 
}

@misc{elizalde2022claplearningaudioconcepts,
      title={CLAP: Learning Audio Concepts From Natural Language Supervision}, 
      author={Benjamin Elizalde and Soham Deshmukh and Mahmoud Al Ismail and Huaming Wang},
      year={2022},
      eprint={2206.04769},
      archivePrefix={arXiv},
      primaryClass={cs.SD},
      url={https://arxiv.org/abs/2206.04769}, 
}

@inproceedings{guzhov2021audioclipextendingclipimage,
      title={AudioCLIP: Extending CLIP to Image, Text and Audio},
      booktitle=icassp,
      author={Andrey Guzhov and Federico Raue and Jörn Hees and Andreas Dengel},
      year={2022},
      url={https://arxiv.org/abs/2106.13043}, 
}

@inproceedings{huang2024multimodaltaskvectorsenable,
      title={Multimodal Task Vectors Enable Many-Shot Multimodal In-Context Learning}, 
      author={Brandon Huang and Chancharik Mitra and Assaf Arbelle and Leonid Karlinsky and Trevor Darrell and Roei Herzig},
      year={2024},
      booktitle=nips,
      url={https://arxiv.org/abs/2406.15334}, 
}

@inproceedings{zhang2024visuallygroundedlanguagemodelsbad,
      title={Why are Visually-Grounded Language Models Bad at Image Classification?},
      booktitle=nips,
      author={Yuhui Zhang and Alyssa Unell and Xiaohan Wang and Dhruba Ghosh and Yuchang Su and Ludwig Schmidt and Serena Yeung-Levy},
      year={2024},
      url={https://arxiv.org/abs/2405.18415}, 
}

@inproceedings{basile2025headpursuitprobingattention,
      title={Head Pursuit: Probing Attention Specialization in Multimodal Transformers},
      booktitle=nips,
      author={Lorenzo Basile and Valentino Maiorca and Diego Doimo and Francesco Locatello and Alberto Cazzaniga},
      year={2025},
      url={https://arxiv.org/abs/2510.21518}, 
}

@inproceedings{jo-myaeng-2020-roles,
    title = "Roles and Utilization of Attention Heads in Transformer-based Neural Language Models",
    author = "Jo, Jae-young  and
      Myaeng, Sung-Hyon",
    editor = "Jurafsky, Dan  and
      Chai, Joyce  and
      Schluter, Natalie  and
      Tetreault, Joel",
    booktitle = "Proceedings of the 58th Annual Meeting of the Association for Computational Linguistics",
    month = jul,
    year = "2020",
    address = "Online",
    publisher = "Association for Computational Linguistics",
    url = "https://aclanthology.org/2020.acl-main.311/",
    doi = "10.18653/v1/2020.acl-main.311",
    pages = "3404--3417"
}

@inproceedings{ma2025cognitivemirrorsexploringdiverse,
      title={Cognitive Mirrors: Exploring the Diverse Functional Roles of Attention Heads in LLM Reasoning}, 
      author={Xueqi Ma and Jun Wang and Yanbei Jiang and Sarah Monazam Erfani and Tongliang Liu and James Bailey},
      booktitle=nips,
      year={2025},
      eprint={2512.10978},
      archivePrefix={arXiv},
      primaryClass={q-bio.NC},
      url={https://arxiv.org/abs/2512.10978}, 
}

@misc{wang2024differentiationspecializationattentionheads,
      title={Differentiation and Specialization of Attention Heads via the Refined Local Learning Coefficient},
      booktitle=2025,
      author={George Wang and Jesse Hoogland and Stan van Wingerden and Zach Furman and Daniel Murfet},
      year={2025},
      url={https://arxiv.org/abs/2410.02984}, 
}

@inproceedings{hojel2024findingvisualtaskvectors,
      title={Finding Visual Task Vectors}, 
      booktitle=eccv,
      author={Alberto Hojel and Yutong Bai and Trevor Darrell and Amir Globerson and Amir Bar},
      year={2024},
      url={https://arxiv.org/abs/2404.05729}, 
}

@inproceedings{todd2024functionvectorslargelanguage,
      title={Function Vectors in Large Language Models},
      booktitle=iclr,
      author={Eric Todd and Millicent L. Li and Arnab Sen Sharma and Aaron Mueller and Byron C. Wallace and David Bau},
      year={2024},
      url={https://arxiv.org/abs/2310.15213}, 
}

@article{qian2025headinformationbottleneck,
author = {Qian, Yukun and Zhuang, Xuyi and Wang, Mingjiang},
year = {2025},
month = {07},
pages = {},
title = {Head information bottleneck (HIB): leveraging information bottleneck for efficient transformer head attribution and pruning},
volume = {2025},
journal = {EURASIP Journal on Audio, Speech, and Music Processing},
doi = {10.1186/s13636-025-00411-8}
}

@misc{zhang2025multiapispoofmultiapidataset,
      title={MultiAPI Spoof: A Multi-API Dataset and Local-Attention Network for Speech Anti-spoofing Detection}, 
      author={Xueping Zhang and Zhenshan Zhang and Yechen Wang and Linxi Li and Liwei Jin and Ming Li},
      year={2025},
      eprint={2512.07352},
      archivePrefix={arXiv},
      primaryClass={cs.SD},
      url={https://arxiv.org/abs/2512.07352}, 
}

@inproceedings{gong2024listenthinkunderstand,
      title={Listen, Think, and Understand},
      booktitle=iclr,
      author={Yuan Gong and Hongyin Luo and Alexander H. Liu and Leonid Karlinsky and James Glass},
      year={2024},
      url={https://arxiv.org/abs/2305.10790}, 
}

@inproceedings{taylor2025improvingaudioclassificationtransitioning,
      title={Improving Audio Classification by Transitioning from Zero- to Few-Shot}, 
      author={James Taylor and Wolfgang Mack},
      year={2025},
      booktitle=interspeech,
      url={https://arxiv.org/abs/2507.20036}, 
}

@inproceedings{chen2019,
author = {Cheng, Kai-Hsiang and Chou, Szu-Yu and Yang, yi-hsuan},
year = {2019},
month = {09},
pages = {1-5},
title = {Multi-label Few-shot Learning for Sound Event Recognition},
booktitle = {2019 IEEE 21st International Workshop on Multimedia Signal Processing (MMSP)},
doi = {10.1109/MMSP.2019.8901732}
}

@misc{brown2020languagemodelsfewshotlearners,
      title={Language Models are Few-Shot Learners}, 
      author={Tom B. Brown and Benjamin Mann and Nick Ryder and Melanie Subbiah and Jared Kaplan and Prafulla Dhariwal and Arvind Neelakantan and Pranav Shyam and Girish Sastry and Amanda Askell and Sandhini Agarwal and Ariel Herbert-Voss and Gretchen Krueger and Tom Henighan and Rewon Child and Aditya Ramesh and Daniel M. Ziegler and Jeffrey Wu and Clemens Winter and Christopher Hesse and Mark Chen and Eric Sigler and Mateusz Litwin and Scott Gray and Benjamin Chess and Jack Clark and Christopher Berner and Sam McCandlish and Alec Radford and Ilya Sutskever and Dario Amodei},
      year={2020},
      eprint={2005.14165},
      archivePrefix={arXiv},
      primaryClass={cs.CL},
      url={https://arxiv.org/abs/2005.14165}, 
}

@inproceedings{Choudhary_2022,
   title={LEAN: Light and Efficient Audio Classification Network},
   url={http://dx.doi.org/10.1109/INDICON56171.2022.10039921},
   DOI={10.1109/indicon56171.2022.10039921},
   booktitle={2022 IEEE 19th India Council International Conference (INDICON)},
   publisher={IEEE},
   author={Choudhary, Shwetank and Karthik, C R and Lakshmi, Punuru Sri and Kumar, Sumit},
   year={2022},
   month=nov, pages={1–6} }

@misc{kumar2023genzgenerativezeroshottext,
      title={Gen-Z: Generative Zero-Shot Text Classification with Contextualized Label Descriptions}, 
      author={Sachin Kumar and Chan Young Park and Yulia Tsvetkov},
      year={2023},
      eprint={2311.07115},
      archivePrefix={arXiv},
      primaryClass={cs.CL},
      url={https://arxiv.org/abs/2311.07115}, 
}

@inproceedings{hendrycks2021measuringmassivemultitasklanguage,
      title={Measuring Massive Multitask Language Understanding},
      booktitle=iclr,
      author={Dan Hendrycks and Collin Burns and Steven Basart and Andy Zou and Mantas Mazeika and Dawn Song and Jacob Steinhardt},
      year={2021},
      url={https://arxiv.org/abs/2009.03300}, 
}

@misc{choi2025exploringfinetuninglargeaudio,
      title={Exploring Fine-Tuning of Large Audio Language Models for Spoken Language Understanding under Limited Speech data}, 
      author={Youngwon Choi and Jaeyoon Jung and Hyeonyu Kim and Huu-Kim Nguyen and Hwayeon Kim},
      year={2025},
      eprint={2509.15389},
      archivePrefix={arXiv},
      primaryClass={cs.SD},
      url={https://arxiv.org/abs/2509.15389}, 
}

@misc{bucher2024finetunedsmallllmsstill,
      title={Fine-Tuned ``Small'' LLMs (Still) Significantly Outperform Zero-Shot Generative AI Models in Text Classification}, 
      author={Martin Juan José Bucher and Marco Martini},
      year={2024},
      eprint={2406.08660},
      archivePrefix={arXiv},
      primaryClass={cs.CL},
      url={https://arxiv.org/abs/2406.08660}, 
}

@inproceedings{ouyang2022traininglanguagemodelsfollow,
      title={Training language models to follow instructions with human feedback},
      booktitle=nips,
      author={Long Ouyang and Jeff Wu and Xu Jiang and Diogo Almeida and Carroll L. Wainwright and Pamela Mishkin and Chong Zhang and Sandhini Agarwal and Katarina Slama and Alex Ray and John Schulman and Jacob Hilton and Fraser Kelton and Luke Miller and Maddie Simens and Amanda Askell and Peter Welinder and Paul Christiano and Jan Leike and Ryan Lowe},
      year={2022},
      url={https://arxiv.org/abs/2203.02155}, 
}

@inproceedings{sun19ttt,
  Author = {Sun, Yu and Wang, Xiaolong and Zhuang, Liu and
  	Miller, John and Hardt, Moritz and Efros, Alexei A.},
  Title = {Test-Time Training with Self-Supervision for Generalization under Distribution Shifts},
  Booktitle = {ICML},
  Year = {2020}
}

@misc{zhang2025aqattrlselfadaptationaudioquestion,
      title={AQA-TTRL: Self-Adaptation in Audio Question Answering with Test-Time Reinforcement Learning}, 
      author={Haoyu Zhang and Jiaxian Guo and Yusuke Iwasawa and Yutaka Matsuo},
      year={2025},
      eprint={2510.05478},
      archivePrefix={arXiv},
      primaryClass={eess.AS},
      url={https://arxiv.org/abs/2510.05478}, 
}

@inproceedings{hendel2023incontextlearningcreatestask,
      title={In-Context Learning Creates Task Vectors},
      booktitle={Findings of EMNLP},
      author={Roee Hendel and Mor Geva and Amir Globerson},
      year={2023},
      url={https://arxiv.org/abs/2310.15916}, 
}

@inproceedings{voita2019analyzingmultiheadselfattentionspecialized,
      title={Analyzing Multi-Head Self-Attention: Specialized Heads Do the Heavy Lifting, the Rest Can Be Pruned},
      booktitle=acl,
      author={Elena Voita and David Talbot and Fedor Moiseev and Rico Sennrich and Ivan Titov},
      year={2019},
      url={https://arxiv.org/abs/1905.09418}, 
}

@misc{clark2019doesbertlookat,
      title={What Does BERT Look At? An Analysis of BERT's Attention}, 
      author={Kevin Clark and Urvashi Khandelwal and Omer Levy and Christopher D. Manning},
      year={2019},
      eprint={1906.04341},
      archivePrefix={arXiv},
      primaryClass={cs.CL},
      url={https://arxiv.org/abs/1906.04341}, 
}

@misc{dogan2024multilabelzeroshotaudioclassification,
      title={Multi-label Zero-Shot Audio Classification with Temporal Attention}, 
      author={Duygu Dogan and Huang Xie and Toni Heittola and Tuomas Virtanen},
      year={2024},
      eprint={2409.00408},
      archivePrefix={arXiv},
      primaryClass={cs.SD},
      url={https://arxiv.org/abs/2409.00408}, 
}

@inproceedings{sridhar2024enhancingtemporalunderstandingaudio,
      title={Enhancing Temporal Understanding in Audio Question Answering for Large Audio Language Models},
      booktitle={NAACL},
      author={Arvind Krishna Sridhar and Yinyi Guo and Erik Visser},
      year={2024},
      url={https://arxiv.org/abs/2409.06223}, 
}

@inproceedings{yang2024uniaudio15largelanguage,
      title={UniAudio 1.5: Large Language Model-driven Audio Codec is A Few-shot Audio Task Learner},
      booktitle=nips,
      author={Dongchao Yang and Haohan Guo and Yuanyuan Wang and Rongjie Huang and Xiang Li and Xu Tan and Xixin Wu and Helen Meng},
      year={2024},
      url={https://arxiv.org/abs/2406.10056}, 
}

@inproceedings{hanif2024palmfewshotpromptlearning,
      title={PALM: Few-Shot Prompt Learning for Audio Language Models},
      booktitle=emnlp,
      author={Asif Hanif and Maha Tufail Agro and Mohammad Areeb Qazi and Hanan Aldarmaki},
      year={2024},
      url={https://arxiv.org/abs/2409.19806}, 
}

@misc{sahoo2025systematicsurveypromptengineering,
      title={A Systematic Survey of Prompt Engineering in Large Language Models: Techniques and Applications}, 
      author={Pranab Sahoo and Ayush Kumar Singh and Sriparna Saha and Vinija Jain and Samrat Mondal and Aman Chadha},
      year={2025},
      eprint={2402.07927},
      archivePrefix={arXiv},
      primaryClass={cs.AI},
      url={https://arxiv.org/abs/2402.07927}, 
}

@misc{olvera2024sounddescriptionexploringprompt,
      title={A sound description: Exploring prompt templates and class descriptions to enhance zero-shot audio classification}, 
      author={Michel Olvera and Paraskevas Stamatiadis and Slim Essid},
      year={2024},
      eprint={2409.13676},
      archivePrefix={arXiv},
      primaryClass={cs.SD},
      url={https://arxiv.org/abs/2409.13676}, 
}

@inproceedings{salinas2024butterflyeffectalteringprompts,
      title={The Butterfly Effect of Altering Prompts: How Small Changes and Jailbreaks Affect Large Language Model Performance},
      booktitle={Findings of ACL},
      author={Abel Salinas and Fred Morstatter},
      year={2024},
      url={https://arxiv.org/abs/2401.03729}, 
}

@misc{deshmukh2024domainadaptationcontrastiveaudiolanguage,
      title={Domain Adaptation for Contrastive Audio-Language Models}, 
      author={Soham Deshmukh and Rita Singh and Bhiksha Raj},
      year={2024},
      eprint={2402.09585},
      archivePrefix={arXiv},
      primaryClass={cs.SD},
      url={https://arxiv.org/abs/2402.09585}, 
}

@misc{belinkov2021probingclassifierspromisesshortcomings,
      title={Probing Classifiers: Promises, Shortcomings, and Advances}, 
      author={Yonatan Belinkov},
      year={2021},
      eprint={2102.12452},
      archivePrefix={arXiv},
      primaryClass={cs.CL},
      url={https://arxiv.org/abs/2102.12452}, 
}

@misc{helber2019eurosatnoveldatasetdeep,
      title={EuroSAT: A Novel Dataset and Deep Learning Benchmark for Land Use and Land Cover Classification}, 
      author={Patrick Helber and Benjamin Bischke and Andreas Dengel and Damian Borth},
      year={2019},
      eprint={1709.00029},
      archivePrefix={arXiv},
      primaryClass={cs.CV},
      url={https://arxiv.org/abs/1709.00029}, 
}

@InProceedings{parkhi12a,
  author       = "Omkar M. Parkhi and Andrea Vedaldi and Andrew Zisserman and C. V. Jawahar",
  title        = "Cats and Dogs",
  booktitle    = "IEEE Conference on Computer Vision and Pattern Recognition",
  year         = "2012",
}

@article{abouelenin2025phi4mini,
  title   = {Phi-4-Mini Technical Report: Compact yet Powerful Multimodal Language Models via Mixture-of-LoRAs},
  author  = {Abdelrahman Abouelenin and others},
  journal = {arXiv preprint arXiv:2503.01743},
  year    = {2025},
  doi     = {10.48550/arXiv.2503.01743},
  url     = {https://arxiv.org/abs/2503.01743}
}



\end{document}